# Long-Term Environmental Stability of Nitrogen-Healed Black Phosphorus

*Valeria S. Marangoni[1,2], Alisson R. Cadore[1], Henrique B. Ribeiro[3], Leandro Hostert, Christiano J. S. de Matos, Cecilia C. C. Silva, Leandro Seixas\*, Camila M. Maroneze\**

MackGraphe – Graphene and Nanomaterials Research Center, Mackenzie Presbyterian University, São Paulo-SP, Postal Code 01302-907, Brazil.

[1] These authors contributed equally to this work.

\* Corresponding authors at: MackGraphe – Graphene and Nanomaterials Research Center, Mackenzie Presbyterian University, São Paulo-SP, Postal Code 01302-907, Brazil.

*E-mail adresses*: camila.maroneze@mackenzie.br (C. M. Maroneze), leandro.seixas@mackenzie.br (L. Seixas)

**Present Addresses**

[2] Valeria S. Marangoni - Centre for Advanced 2D Materials (CA2DM), Faculty of Science, National University of Singapore (NUS), 117546 Singapore

[3] Henrique B. Ribeiro - Department of Applied Physics, Stanford University, Stanford, CA

**Keywords.** Black phosphorus, environmental stability, defects, nitrogen-healing, e-beam irradiation

**Abstract**

The unique optoelectronic properties of black phosphorus (BP) have triggered great interest in its applications in areas not fulfilled by other layered materials (LMs). However, its poor stability (fast degradation, i.e. <<1 h for monolayers) under ambient conditions restricts its practical application. We demonstrate here, by an experimental-theoretical approach, that the incorporation of nitrogen molecules ($N_2$) into the BP structure results in a relevant improvement of its stability in air, up to 8 days without optical degradation signs. Our strategy involves the generation of defects (phosphorus vacancies) by electron-beam irradiation, followed by their healing with $N_2$ molecules. As an additional route, $N_2$ plasma treatment is presented as an alternative for large area application. Our first principles calculations elucidate the mechanisms involved in the nitrogen incorporation as well as on the stabilization of the modified BP, which corroborates with our experimental observations. This stabilization approach can be applied in the processing of BP, allowing for its use in environmentally stable van der Waals heterostructures with other LMs as well as in optoelectronic and wearable devices.

Ultra-thin black phosphorus (BP) flakes have been extensively studied owing to their unique optoelectronic properties,[1–4] highlighting the strong light-matter interaction in a broad frequency range (from mid-infrared to visible),[5] which allows important applications in the areas of optical sensing and detection[6–8], as well as in optical communications.[9] The tunable direct bandgap, from 0.3 eV for bulk to up to 1.7 eV for monolayers,[10–12] the high charge carrier mobility (up to $\sim 10^4$ cm$^2$V$^{-1}$s$^{-1}$ for monolayers),[13] and the high $I_{on}/I_{off}$ current ratio in field effect transistors (up to $\sim 1\times 10^7$)[14] provide BP with relevant advantage over other layered materials (LMs) like graphene (zero bandgap material) and transition-metal dichalcogenides (TMDs - which show lower carrier mobilities, up to $\sim 10^3$ cm$^2$V$^{-1}$s$^{-1}$, when compared to BP devices).[15,16] Despite the exciting possibilities already identified for optoelectronic devices, a major challenge that still needs to be overcome is the high instability of BP, that rapidly degrades under ambient conditions (i.e. <<1h for monolayers).[17–22] It is known that the reaction with oxygen initially leads to the formation of P-O bonds, resulting in $P_xO_y$ oxide species, which further react with water molecules, culminating with the formation of phosphoric acid on the BP surface and the complete disintegration of the material.[23–27]

Physical and chemical approaches have been used both to improve the BP stability in air and modulate its properties.[28,29] The coating with $Al_2O_3$ by atomic layer deposition[22,30,31] or other LMs, such as hexagonal boron nitride (hBN) and graphene,[32–35] are examples of physical passivation. The chemical functionalization is also an efficient way to protect BP and to manipulate its chemical and electronic properties.[21,36,37] The modification of the BP surface with aryl diazonium molecules, for example, was ascribed to suppress its degradation but induced a simultaneous p-type doping effect.[38] Non-covalent modification of BP with anthraquinone molecules minimized its degradation and added new features such as redox functionality and additional charge storage capacity.[39] Ionic liquids have also been used to quench reactive oxygen species and stabilize BP,[40] while the fluorination has shown to improve its long term air-stability by repelling the $O_2$ due to the highly electronegative fluorine adatom, with changes in the optical, chemical, and structural properties of BP.[41] *In situ* surface functionalization of BP with the alkali metal potassium (K), an excellent electron donor, enhance its electron transport and modify its bandgap, allowing a wide tunability of its optoelectronic properties.[42,43] However, due to the high reactivity of K in air, the functionalization did not improve the BP stability.[42,43] Transition metal-phosphorus complexes can also reduce the BP oxidation by the coordination of the lone pair of electrons with the empty orbitals of the metal atoms.[44] Although much progress has been achieved, some key problems still need to be addressed. In the encapsulation method, for example, the inevitable trap of defects, impurities, water and oxygen can compromise the BP performance, and the covalent functionalization can reduce its carrier mobility.[35,36] Besides that, it is important to highlight that both methods, do not allow the development of heterostructures based on the BP nanosheets.[45,46]

A less explored approach is the substitutional doping into the crystal lattice of BP. A theoretical study has compared the stability of phosphorus allotropes (black and blue phosphorus) after substitutional doping with different elements.[47] They revealed that N, O, F, Si, and S could improve BP stability, and highlighted nitrogen as the best dopant among them. Nevertheless, and as far as we know, this effect has not been experimentally demonstrated. So far, Valappil and co-workers[48] have tried to modify electrochemically obtained BP quantum dots with nitrogen-containing molecules; however, the experiments were performed by a wet-process in a nitrogen-rich electrolytes (tetraethylammonium tetrafluoro-borate) and organic solvents, and the results do not show considerable BP stability improvement. Therefore, the demonstration of nitrogen incorporation in BP crystals in a fully controllable and dry way and demonstrating long BP stability is still lacking. It is worth noting that recently Ji and co-workers[49] have studied the synthesis of metastable black phosphorus–structured nitrogen under extreme conditions of pressure and temperature (146 GPa and 2200 K), and observed that molecular nitrogen was transformed into extended single-bonded structure with strong anisotropy.

Herein, by an experimental-theoretical approach, we demonstrate how the feasible incorporation of nitrogen on the BP structure (N-BP) can improve its stability in ambient conditions. We show that treated N-BP samples are stable for at least 8 days in air while non-treated samples show severe degradation features before the first day. The proposed strategy is based on a controlled electron beam (e-beam) irradiation of the BP surface, followed by the healing of the as-created vacancies with $N_2$ molecules. This approach provides high spatial resolution and density control of the nitrogen doping level. We also demonstrate a simpler way to improve BP stability by using $N_2$ plasma, which is a good candidate for large area treatments. The possible mechanisms involved both in the nitrogen healing and in the stabilization of the modified N-BP are investigated by first principles calculations based on density functional theory (DFT), and the computational results are consistent with our experimental findings.

A schematic representation of the procedure used to modify the BP flakes is shown in **Figure 1(a-c)**. The samples are prepared by mechanical exfoliating bulk BP on a 300 nm $SiO_2$/Si substrate. Then, they are exposed to the e-beam in a scanning electron microscope (SEM), **Figure 1(a)**. In our experiments, we have changed the e-beam energy as well as the exposure time, i.e. the time over which the e-beam impinges on the selected sample (see Methods for details). After exposing the selected flake region with the pre-defined energy and time, the sample is transferred to the pre-chamber of the SEM (in vacuum) and then ventilated with anhydrous $N_2$ before its removal from the SEM. The e-beam is expected to create defects[50] in BP structure, **Figure 1(b)**, which can be subsequently healed with $N_2$ molecules during the ventilation step, **Figure 1(c)**.

**Figure 1(d)** shows the SEM image of a region with several exfoliated BP flakes on the $SiO_2$ substrate, which was briefly exposed to the e-beam (1 kV and magnification of 750x) only for the image acquisition (< 30 s). The highlighted region (marked area in **Figure 1(d)**) corresponds to a BP flake, **Figure 1(e)**, exposed to the e-beam (1 kV) for a selected period of 5 min at a higher magnification (7500x), after the image acquisition (< 30 s). After the removal from the SEM, the sample is then exposed to ambient conditions (19±1 ºC and humidity of 60±5%), analyzed by atomic force microscopy (AFM) and monitored by optical microscopy and Raman spectroscopy (See Methods Section for details). The thickness of the exposed flake is determined by AFM immediately after the exposure to the e-beam, **Figure 1(f)**. The line profile (inset) indicates two different thickness regions (~17 and 31 nm). We then characterize our irradiated BP samples by Raman spectroscopy[26,51–55] (See Methods Section for details) to further confirm the structural integrity of our material before and after the $N_2$ healing. Raman spectra of the N-BP flake are acquired at the select points on the sample indicated in the AFM image (**Figure 1(f)**) after 24 hours under environmental conditions following the e-beam exposure. As can be observed in **Figure 1(g)** and **Table S1**, the vibrational modes of the modified N-BP present high similarity with those of the pristine BP flake, which is a reference flake that was neither exposed to the e-beam nor to ambient conditions, with the typical $A_g^1, A_g^2,$ and $B_{2g}$ modes, indicating no significant structural modification after the e-beam exposure.[51,52] The use of higher e-beam energy (15 kV), as proposed by Goyal and co-workers[50], has shown to induce structural modification to the BP Raman modes. Therefore, we have kept the energy at lower levels in which we do not detect changes in the BP Raman modes induced by the e-beam irradiation **(Figure S1)**.

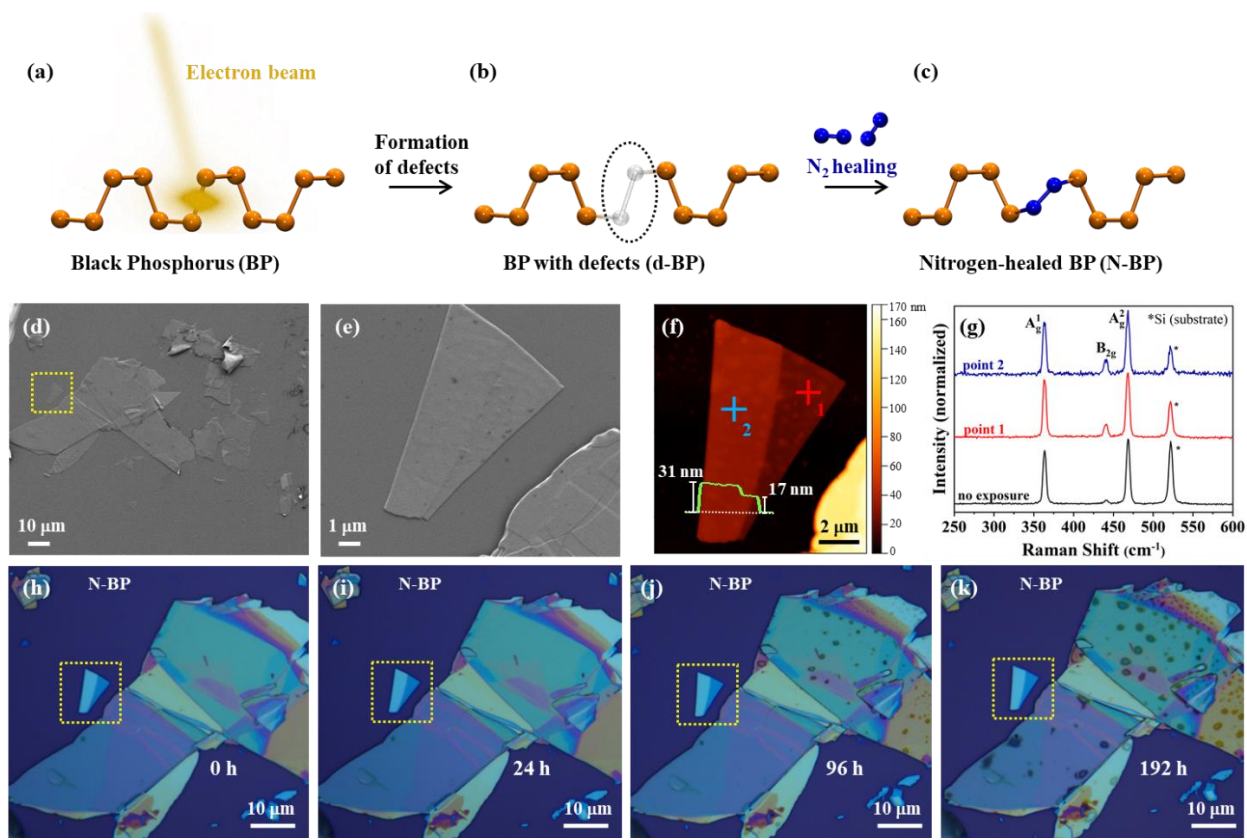

**Figure 1. Atom level representation and BP flake monitoring after e-beam exposure and $N_2$ healing**. (a-b) Schematic representation of the generation of defects in BP after exposure to the e-beam, followed by the (c) healing with nitrogen. (d-e) SEM images of the BP flakes (marked area) exposed to the e-beam. In both cases we use 1 kV: in (d), the flakes are exposed for < 30 s at 750x, while in (e) the magnification is 7500x and the selected flake is exposed for 5 min. (f) AFM image of the same flake after the e-beam exposure and $N_2$ healing. The inset shows the height profile. (g) Raman spectra of the exposed N-BP flake obtained at the indicated points (red and blue crosses in the AFM image) with different thicknesses after 24 hours under environmental conditions. A spectrum for a non-exposed flake with similar thickness is also presented for comparison (black curve). (h-k) Optical microscope images after the exposure of the sample to ambient conditions for different periods of time.

The sample stability over time was then evaluated by optical microscopy, **Figure 1(h-k)**. The formation of bubbles, typically observed in degraded ultrathin BP due to oxygen and humidity,[17–23] is not observed for the healed flake even after 192 h (8 days), unlike other non-treated regions of the sample which clearly show the appearance of bubbles after 96 h. It has been demonstrated that for pristine BP flakes the degradation begins just after the exfoliation,[17–22] and after 96 h the flakes with thickness of ~27 nm already present a high density of bubbles on the surface.[56] After 196 h, these bubbles are much larger by coalescence and normally deteriorates the BP etching away the thinner parts of the flakes. Nonetheless, it is worth noting that, according to our observation, the flakes that were exposed to the e-beam for a short period (imaging only, < 30 s) also undergo a slower degradation than the flakes that were not exposed to the e-beam at all. Under the same ambient conditions, non-treated BP flakes have the surface highly covered by bubbles in less than 24 h (**Figure S2**).

Raman spectroscopy has been extensively used to study defects and degradation of LMs such as graphene, hBN, and TMDs.[51,53,57–59] So far, however, no universal Raman signatures have been assigned to BP degradation. For instance, in previous works lower intensities in the Raman shift spectra, lower $A_g^1/A_g^2$ intensity ratios, and the emergence of broad features between the $A_g^2$, and $B_{2g}$ vibrational modes have all been identified as signatures of the oxidation and degradation, as well as a shift in the peak position.[23,26,27,29,34,50,60,61] However, in our analysis (data not shown), we do not see a clear Raman signature for degradation in our samples, suggesting that Raman spectroscopy may not be the best tool to characterize the BP degradation. Therefore, we evaluate the environmental stability by AFM, as previous works have shown that the oxidation increases the BP surface roughness[31,50] and crystal volume.[18] For this experiment, the sample is exposed for 10 min at 5 kV with 10000x (selected area in **Figure 2(a)**, while the overall SEM image is exposed for < 30 s). Before acquiring the AFM image, **Figure 2(b)**, the sample is exposed to environmental conditions for 72 h. **Figure 2(c)** shows an AFM height profile obtained as indicated in **Figure 2(b)**. The AFM image clearly shows the enhanced stability of the region that has been selectively exposed to the e-beam (marked area in **Figure 2(a)**). The root mean square (RMS) roughness changes from about 2.26 to 1.18 nm (areas indicated in **Figure S3b)** as we go from a non-treated (i) to exposed (ii) regions. In the area where the degradation of BP is more accentuated (region "i"), in addition to higher bubble density, the height profile, **Figure 2(c)**, indicates the formation of a rougher surface, as well as an increase to the profile of the height. This observation (increase in volume) is in agreement with previous AFM findings, and was attributed to water absorption on BP over time.[18] The stability over time of the same flake stored under ambient conditions is shown in **Figure S3(a)**. Note that the enhanced stability of a BP flake after e-beam exposure has also been reported experimentally by Goyal and co-workers.[50] However, it is worth mentioning that in that work the authors have assigned such stability to strong structural changes of the BP induced by a high e-beam energy (15 kV) and to a reduction in the formation of $PO_x$ after e-beam irradiation. Nevertheless, they do not show any theoretical prediction to support their claims and do not investigate or consider the influence of external molecules which may bind to the BP structure.[47] In our study, we have tested different e-beam acceleration voltages and exposure times on BP flakes of different thicknesses. Although our analysis suggests that the $N_2$ healing occurs in all cases, reducing the degradation process, the dependence of environmental stability with thickness, irradiation time, and energy is not clear at this stage and needs further investigation. In the Supplementary Information, we present optical images (**Figures S3, S4, and S5**) of different flakes exposed to distinct e-beam acceleration indicating the long environmental stability of samples treated with different energies and exposure times. Therefore, our experimental data show strong signs that e-beam irradiation and nitrogen healing significantly improve the environmental stability of BP.

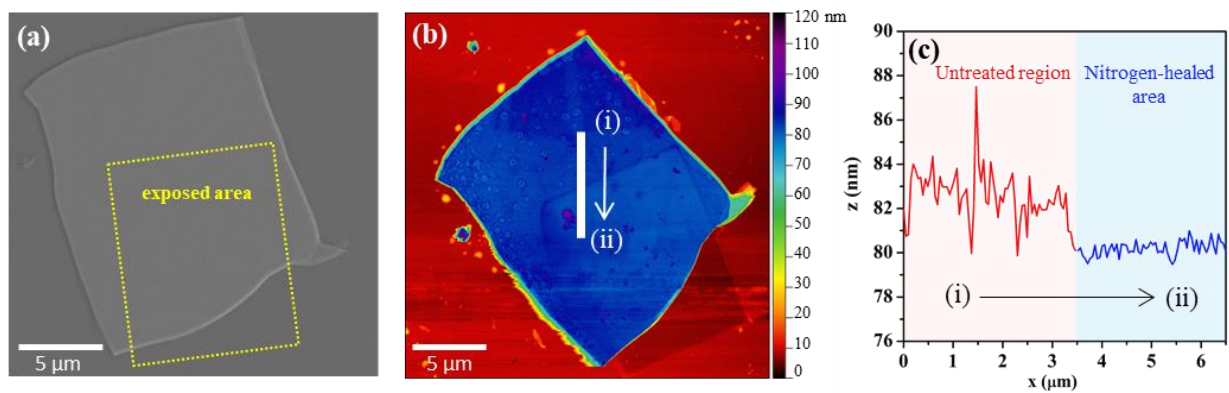

**Figure 2. Topological and structural analysis of BP flakes exposed to the e-beam.** (a) SEM image of a BP flake exposed to the e-beam, where the highlighted area represents the region exposed at 5 kV, magnification of 10000x for 10 min. (b) AFM image and (c) the respective height profile for the same flake after 72 hours of exposure to ambient.

The high spatial resolution of the e-beam treatment system allows us to investigate the nitrogen healing effect in the same flake, which is important to guarantee the direct comparison with the initial properties. However, the small processing areas and the difficulty to locate thinner flakes in the SEM can limit the investigation and future applications. Herein, $N_2$ plasma is also proposed and used as an alternative route to treat large areas. For the experimental details of the plasma process see Methods Section. $N_2$ plasma-treated sample is compared to a non-treated sample (control) and Ar plasma-treated BP as shown in **Figure 3**. From the optical images, the Ar plasma-treated BP (d-BP), **Figure 3(d-f)**, shows the fastest degradation rate, probably due to the significant formation of defects,[62,63] which undergo intense oxidation and degradation after exposure to ambient conditions. After 48 h under ambient environment, the control flakes present a highly oxidized surface, which is covered by bubbles, **Figure 3(a-c)**, while the Ar plasma-treated BP is already completely degraded, with no flakes remaining on the substrate, **Figure 3(d-f)**. In stark contrast, $N_2$ plasma-treated BP shows an evident improvement in the stability, and the material is mostly preserved after the same exposure period, **Figure 3(g-i)** and **Figure S6**. It is worth noticing that further experiments are still necessary to determine the optimized conditions for $N_2$ plasma processing. However, our results are a strong evidence that, during the plasma treatment, nitrogen species are incorporated into the BP lattice simultaneously with the removal of the P atoms, resulting in a modified material with enhanced environmental stability. Ar [62,63] and $O_2$ [30] plasma are often used to etch BP. The $O_2$-etched BP shows higher stability than the Ar-etched material, which has been attributed to the formation of a protective $P_xO_y$ encapsulation layer.[64] So far to the best of our knowledge, however, exposure of BP to $N_2$ plasma has not been evaluated.

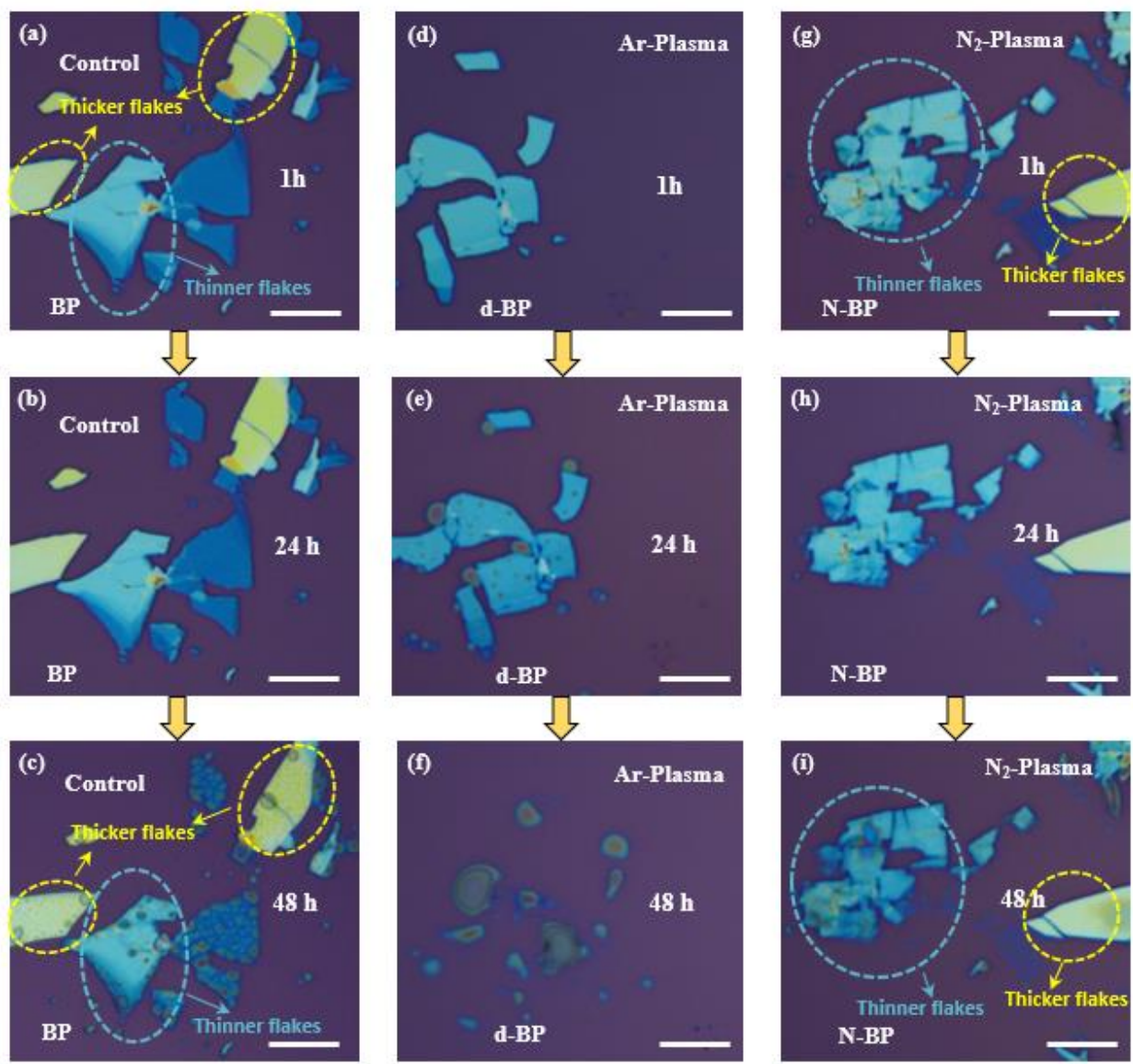

**Figure 3. Plasma treatment of BP flakes.** Optical microscopy images, taken 1, 24 and 48 hours in ambient conditions after sample preparation, for pristine BP (a-c), BP exposed to Argon plasma for 5 min (d-f), and BP exposed to $N_2$ plasma (N-BP) for 5 min (g-i). The highlighted areas (blue and yellow circles) show regions with similar optical contrast (thickness) in the control and $N_2$-plasma treated samples.

To further investigate the nitrogen incorporation into BP crystals in the plasma-treated samples, we have performed an X-ray photoelectron spectroscopy (XPS) investigation. **Figure 4(a)** shows the N1s and P2p high-resolution spectra for the non-exposed (control) and $N_2$ treated

BP on Si/SiO$_2$ substrates. The elemental P peaks can be observed at ~130 eV, [65] while the peak ~135 eV is attributed to oxides species (P-O), where P$_2$O$_5$ is suggested to be the most stable form.[29] The broad peak around ~127 eV originates from the *s* photoelectron Si satellite from the substrate.[22] For the control sample, we can observe a relative decrease in intensity of the elemental P peaks, followed by an increase in the relative amount of oxides species. This is expected due to the inevitable air exposure of the samples before the measurements, since they are transported in a vacuum desiccator but exposed to environmental conditions during sample preparation.[66] Under the same conditions, the N$_2$-treated BP still presents well defined and relatively intense peaks from the BP structure at ~ 130 eV. By comparing the high-resolution N1s spectra, we clearly see differences between the two systems, for instance, N$_2$-treated BP presents more, well defined, and higher intensity peaks. Also, the increase in the FWHM of the elemental P peaks also agrees with the higher degree of oxidation in the control sample (**Table S2**).[22] Since the XPS spot size is ~400 μm and the mechanical exfoliated material do not cover the whole substrate surface, we have also analyzed the N1s region of the Si/SiO$_2$ substrate before and after the N$_2$ plasma treatment under the same conditions (**Figure S7**). The results confirm the modification of the SiO$_2$ surface with nitrogen groups,[67] and, therefore, the two peaks around 403.4 and 399.1 eV in the N$_2$-treated samples are from the substrate. Nevertheless, the additional peaks at 401.9 and 400.4 eV observed in **Figure 4(a)** for the N$_2$ plasma-treated samples should be from the modified N-BP crystal. Wang and co-workers have attributed the peaks at about 398 eV and 400 eV to the nitrogen bound to two and three phosphorus atoms, respectively.[68] Analogously to the e-beam exposed samples, the Raman spectra of the N$_2$-plasma treated samples present no significant difference in the Raman modes and FWHM as can be seen in **Figure 4(b)** and **Table S3. Figure 4(b)** also shows the Raman spectra for the control sample (after exfoliation and 48h under the environment) and the data is in stark contrast with the optical images shown in **Figure 3(a-c)**, reinforcing that Raman spectroscopy is not the ideal tool to investigate the BP degradation.

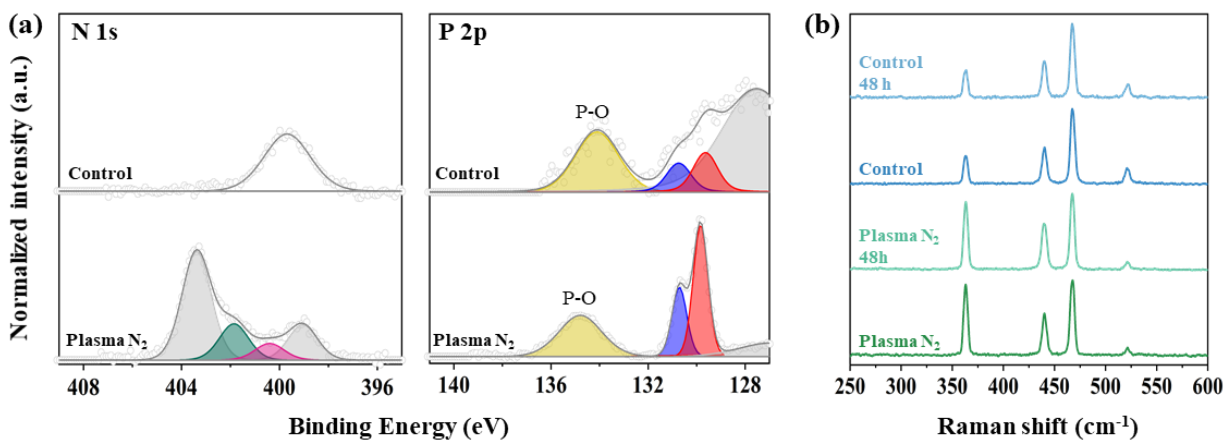

**Figure 4. Effect of the $N_2$ plasma treatment on BP flakes.** (a) High resolution N1 and P2p XPS spectra and (b) Raman spectra for pristine BP (control) and exposed $N_2$-plasma samples.

The presence of a wide variety of natural defects is also observed in BP crystals, and they are easily formed when compared to those in graphene.[69] Riffle and co-workers showed by scanning tunneling microscopy and spectroscopy (STM/S) that commercial BP bulk crystals have a significant amount of native defects (P vacancies), explaining the p-type doping usually observed in BP.[70] A previous theoretical study demonstrated that the vacancies in BP have a strong affinity to oxygen, with an oxidation rate up to 5000x faster than the perfect BP lattice site.[71] Based on that, we perform the mechanical exfoliation of BP (using the same bulk crystal) in two different glove-box systems under $N_2$ and Ar atmosphere, followed by storage of both samples under the same ambient conditions (see Methods section for details). Confirming our expectations, the stability of the flakes exfoliated under the $N_2$ atmosphere is higher than that carried out in the Ar environment, **Figure S8**, suggesting that a simple $N_2$ environment may also have an important effect on the healing of the native defects. After 48 h of exposure to ambient oxygen and humidity, the degradation of the BP surface (size and number of bubbles) is distinctly higher in the sample exfoliated in Ar atmosphere. Therefore, considering all three different approaches discussed above, we have shown that BP samples treated under nitrogen environment present higher environmental stability when compared to both pristine and Ar plasma treated samples.

To gain further insight concerning the formation of defects on the BP structure and the subsequent healing with nitrogen, we have performed DFT simulations. An earlier theoretical study has evaluated the creation of defects in a single phosphorene layer under an e-beam.[72] Depending on the magnitude and direction of the collision, vacant sites with subsequent recombination, sputtering or displacement of P atoms to distant adatom sites can occur.[72] Another possibility is the formation of single vacancies, which can be easily expanded since the P atoms in the vicinity of the vacancies are likely less tightly bound than those in the pristine lattice.[72] Thus, the exposure time and e-beam energy may influence the type and density of defects on the BP surface, which corroborates our observation that the stabilization is dependent on the e-beam energy, SEM magnification (energy dose per area), and the duration of the exposure to the e-beam. The most stable native point defects in BP are reconstructions of phosphorus divacancies.[73] Since the samples are submitted to an $N_2$ atmosphere at the specimen pre-chamber just after the e-beam exposure, we propose that the $N_2$ molecules can heal the divacancy-related defects generated by the e-beam, resulting in a system with improved stability.

To support our experimental claims, we calculate the formation energies ($\Delta E_f$) for three different reconstructions of phosphorus divacancies, **Figure 5(a)**, labelled here DV-(5|8|5)-1, DV-(4|10|4), and DV-(5|8|5)-2, whose values are respectively 1.50, 2.27 and 3.20 eV. The numerical indices in these labels represent the number of sides of the polygons formed by P-P bonds in the defect reconstruction. All three defects have very low formation energies, much smaller than the divacancy formation energy in graphene (~7 eV).[74] To evaluate one possible mechanism for the $N_2$ healing, the energy barriers are also calculated with the nudged elastic band (NEB) method (See Methods section for details). It starts with an $N_2$ molecule physiosorbed on the divacancy defect and ends up with the $N_2$ replacing the missing P atoms. For the DV-(4|10|4) and DV-(5|8|5)-2 defects, the incorporated $N_2$ are approximately orthogonal to the basal plane of BP (N-BP$_{ortho}$), whereas for DV-(5|8|5)-1 the $N_2$ are parallel to the basal plane (N-BP$_{para}$), **Figure S9**. The kinetics of this reaction is related to the transition state energy ($E_{TS}$), which has the lowest value for the DV-(5|8|5)-2 defect, $E_{TS}$ = 0.99 eV, **Figure 5(b)**. The difference in energy between the final state and the initial state is negative ($E_{FS}$ = -0.38 eV) for this defect. For the $N_2$ incorporation in DV-(4|10|4) and DV-(5|8|5)-1, the transition state energies were 2.60 eV and 1.92 eV, and the final state energies were 1.48 eV and 0.53 eV, respectively (with the initial state energies set to zero in these studies). It is also observed that the final state and transition state energies for the $N_2$ incorporation are correlated with the formation energies of the defects, as depicted in **Figure 5(c)**. Since the $N_2$ incorporation approximately restores the crystal lattice of BP, we call this process $N_2$ healing of BP. Snapshots of the initial, transition and final states for the $N_2$ healing in the DV-(5|8|5)-2 defect are shown in **Figure 5(d)**. It is important to mention that some atomic restructuring of the nitrogen-healed BP was also evaluated considering the possibility of the nitrogen atoms not being adjacent, but with a phosphorus atom in between (**Figure S10**). The results have shown that such configuration is also feasible and 1.08 eV lower in energy than the configuration with adjacent nitrogen atoms.

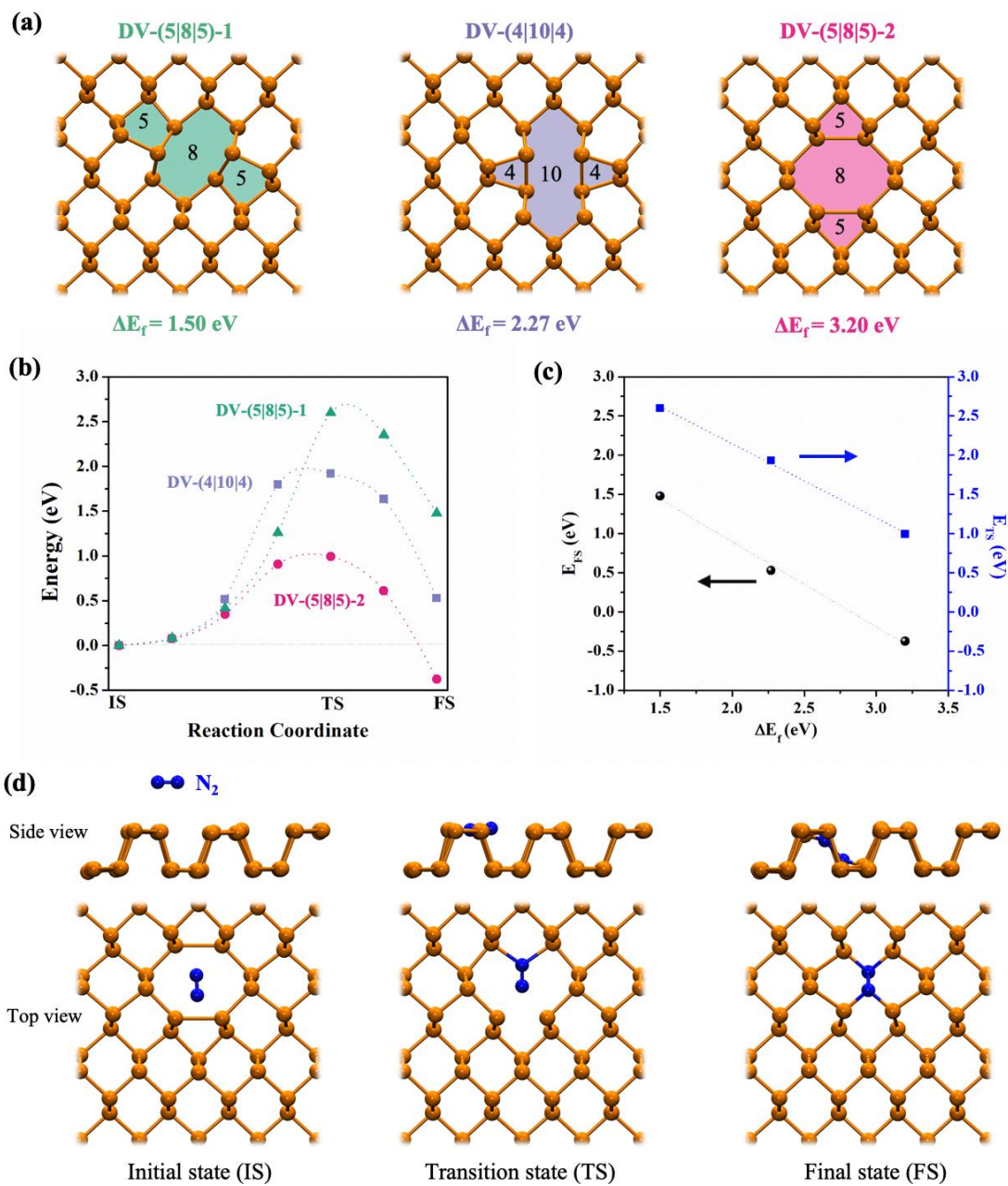

**Figure 5. Mechanism of nitrogen healing of BP.** (a) Ball-and-stick representation of three different reconstructions of P divacancies: DV-(5|8|5)-1, DV-(4|10|4), DV-(5|8|5|)-2. (b) Energy barrier profiles for nitrogen healing under different P divacancy reconstructions. (c) Linear dependence of the transition state (TS) and final state (FS) energies of the $N_2$ healing process with the P divacancy formation energy. (d) Side and top views of three snapshots (initial, transition and final states) of the $N_2$ healing of BP in a DV-(5|8|5)-2 defect.

We have also calculated the electronic properties of a pristine BP monolayer, as well as of the divacancy DV-(5|8|5)-2 monolayer (d-BP), and an $N_2$-healed monolayer (N-BP), as shown in **Figure 6(a-c)**. The pristine BP shows semiconducting properties with a direct bandgap of 0.94 eV. This bandgap is underestimated due to the exchange-correlation functional used in our density functional calculations. Further calculations with hybrid exchange-correlation functional (HSE06 functional) increases the band gap of the BP monolayer up to 1.70 eV.[75] In our samples, the experimental bandgap should be much smaller due to the thickness effect.[10–12] For the BP with DV-(5|8|5)-2 defect (d-BP), as depicted in **Figure 6(b)**, there are two unoccupied localized states between the valence and the conduction bands. These localized states are quite reactive and favourable to the incorporation of nitrogen into the structure. **Figure 6(c)** presents the electronic band structure of the N-BP$_{ortho}$ configuration. The nitrogen healing extinguishes the localized states and restores the band structure to a profile very similar to the pristine BP. For these band structures, the orbital compositions of each state are also calculated and are plotted as a coloured map. Nitrogen contributions are shown in blue, while phosphorus contributions are shown in red. We also calculate the projected density of states (PDOS) normalized by the number of atoms shown in **Figure 6(a-c)**. We can observe that the contributions from nitrogen in N-BP form a small peak in the valence band. **Figure 6(d-f)** shows the energies calculated for the oxidation process of BP, d-BP and N-BP. It can be clearly seen that N-BP (-0.70 eV) is less prone to oxidation than the pristine (-1.69 eV) or the defective material (-2.72 eV). These increases in formation energies should result in a proportional increase in energy barriers for oxidation processes, and in the slowing of the kinetics of the oxidation reaction. For the N-BP$_{ortho}$, we calculate the space-resolved local density of states (LDOS) for the conduction band, from CBM to CBM + 1.0 eV; and for the valence band, from VBM – 1.0 eV to VBM, as shown in **Figure 6(g-h)**. Although the band

structures of the pristine BP and the N-BP are very similar, the orbital contributions of the N cause a different spatial resolution from the pristine material. The contribution of nitrogen orbitals in the Bloch states to the conduction band is delocalized in the zigzag direction, and localized in the armchair direction, forming a quasi-one-dimensional state as displayed in **Figure 6(g)**. For the valence band, the contribution of the N orbitals is localized close to the defect, with a small increase in the density of states of P atoms close to the $N_2$ defect, as shown in **Figure 6(h)**. Finally, it is worth mentioning that the electronic properties of the pristine BP are not expected to change in N-healed BP, at least for doping levels up to 6.25%, including the anisotropy and the thickness-controlled band gap.[47] Therefore, we foresee that our approach can be easily implemented as an additional step in the fabrication process of different devices such as BP field-effect transistors and photodectors.[6–8] Furthermore, we believe that both electrical and optical properties of N-doped BP flakes can be probed and studied in more details in future works, for accessing the penetration depth of the e-beam (and $N_2$ healing) and for estimating the degree of N replacement in the BP structure.

In summary, we demonstrated, by theory and experiments, that the incorporation of nitrogen into the structure of black phosphorus improves its stability under ambient conditions. We showed that treated N-BP samples are stable for at least 8 days in air while non-treated samples show severe degradation features before the end of the first day. The proposed mechanism involves the formation of defects (phosphorus vacancies) in the BP structure, followed by the healing of the vacancies with nitrogen molecules. The nitrogen substitution, controlled by e-beam irradiation, presents high spatial resolution, allowing its control at nanometric scale. We also demonstrated the use of $N_2$ plasma irradiation as an alternative route to scale up the process.  Our contribution provides a deeper understanding concerning the strategies to stabilize the BP and paves the way towards their future use as stable electrodes, thin films, heterostructures and optoelectronic devices.

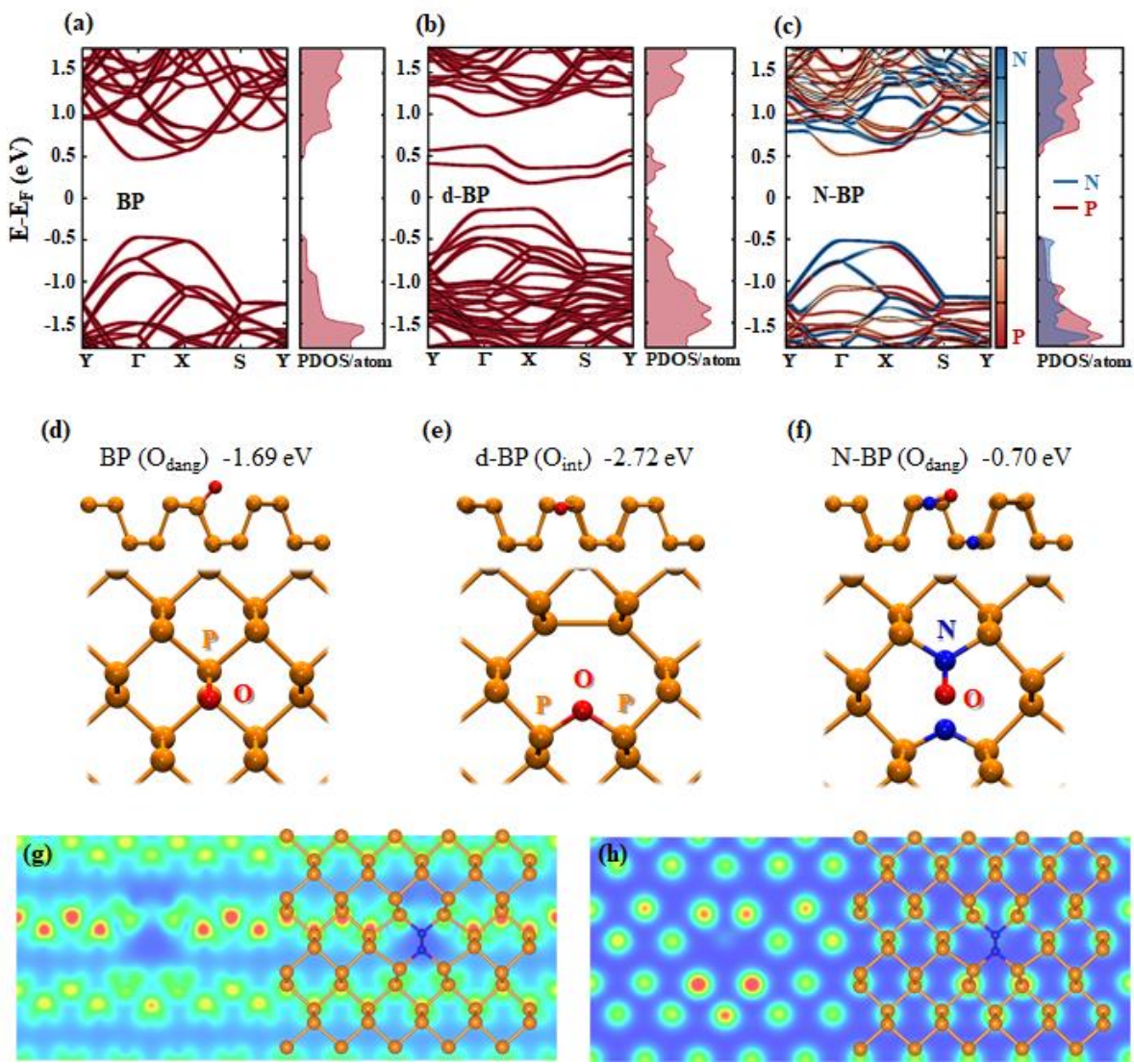

**Figure 6. Electronic properties of BP, d-BP and N-BP.** Band structure and density of states for: (a) pristine BP monolayer, (b) BP with divacancy defect (d-BP), (c) $N_2$ healed BP (N-BP). Oxidation of BP, d-BP and N-BP: ball-and-stick representation of oxygen adsorption on (d) pristine BP, (e) d-BP, (f) N-BP. (g) LDOS for N-BP conduction band (CBM < E < CBM + 1.0 eV). (h) LDOS for N-BP valence band (VBM – 1.0 eV < E < VBM). The LDOS is shown as a volume slice plot with colormap with Blue-Green-Red palette for values from 0 to 0.015 a.u. High concentrations of electronic states are shown in red.

**Methods:**

*Sample preparation and e-beam nitrogen healing*: Black Phosphorus (BP) flakes were prepared by micromechanical exfoliation of bulk BP (HQ Graphene) on 300 nm thick $SiO_2$/Si substrates using the Scotch tape method in a glove box under $N_2$ environment (< 3% of $O_2$). The substrates were placed on the SEM stubs and packed in sealed containers inside the glovebox. Then, they were removed from the glove box and quickly transferred to the Scanning Electron Microscope (SEM) pre-chamber. The estimated period of air exposure (temperature: 19±1ºC and humidity: 60±5%) was around 120 seconds. The exposure to the electron beam was performed using a JEOL JSM-7800F - SEM operated at ~$10^{-7}$ Torr. Different parameters, such as electron beam voltage, magnification and exposure time were evaluated. We have tested 1 kV, 5 kV, and 10 kV as acceleration voltage for the e-beam, as well as 1 min, 5 min, and 10 min for the continuous exposure. The magnification used were 750x for the overall image and 7500x and 10000x for the e-beam irradiation. After the e-beam treatment, the substrates were transferred from the analysis chamber to the pre-chamber (under vacuum) and ventilated with ultra-pure and anhydrous $N_2$ gas. After removal from the SEM, the samples were kept on Petri dishes in a clean room (ISO 6) under ambient conditions (temperature: 19±1ºC e humidity: 60±5%) without any additional protection. The control experiments were carried out with BP flakes manipulated under the same conditions, except for the irradiation and healing procedures.

*Sample characterization:* Optical images were obtained using Nikon and Olympus optical microscopes. Atomic force microscopy (AFM) measurements were performed using a Bruker Dimension Icon Microscope operated in ScanAsyst tapping mode and scan lines of 512 under ambient conditions. Confocal Raman spectroscopy was carried out in a confocal Raman microscope WITec Alpha 300R, with excitation wavelength of 532 nm and a 100x objective with a numeric aperture of 0.9. To guarantee that each Raman spectrum was recorded under the same conditions of light polarization, an analyzer was placed parallel with the polarization of the incident light. By rotating the sample and minimizing the $B_{2g}$ Raman signal, we guarantee that the sample is always oriented with one of the main crystallographic axes along the light polarization axis.[55] After oriented, a small rotation (<10º) was performed in order to observe and analyze the $B_{2g}$ BP Raman mode. The laser power was kept constant at 100 μW during all measurements to avoid sample degradation. X-ray Photoelectron Spectroscopy (XPS) spectra were obtained using the X-ray Photoelectron Spectrometer K-Alpha (Thermo Scientific) with Al Kα micro-focused monochromator x-ray source and spot size of 400 μm, total acquisition time of 7 min 32.5 seconds, 50 scans, energy step size of 0.1 eV, analyzer mode CAE with 50 eV of pass energy and number of energy steps of 181. Shirley type background and peak fitting were performed by means of CasaXPS software (version 2.1.19). The peaks were fitted using Gaussian-Lorentzian GL(30) line shape.[76] The spectra were normalized with respect to the highest peak intensity for ease of comparison.

*BP exfoliation in controlled atmosphere and plasma treatment:* To further explore the role of the nitrogen in the stabilization of the BP flakes, the incorporation of nitrogen in two different conditions were also evaluated. In the first case, bulk BP from the same crystal (HQ Graphene) was placed in two different glove boxes (Ar and $N_2$ atmosphere). They were mechanically exfoliated using the same procedures previously described. Then, they were kept under the respective atmosphere for 1 week (Ar or $N_2$) followed by the exposure to the ambient conditions. For the plasma treatments, BP crystals were mechanically exfoliated as previously described under $N_2$ environment. The substrates were removed from the glove box and quickly (<1min) transferred to the plasma system (SPI Supplies - Plasma Prep III Etcher). The system chamber was evacuated (120 mTorr) and either the $N_2$ or the Ar gas line was opened until the pressure reached 255 mTorr. The flakes were submitted to the respective plasma (25 W) for 5 min, and then stored under ambient conditions (temperature: 19±1ºC e humidity: 60±5%) without any additional protection.

*Ab initio simulations*: We carried out *ab initio* simulations based on density functional theory with localized atomic basis implemented in Siesta code.[77] The basis set for Kohn-sham orbitals was defined with DZP functions, and energy shift of 0.030 eV. We set mesh cutoff of 300 Ry, supercells with 5x3x1 unit cells and k-points with the Gamma-centered grid of 3x5x1 for supercells. The exchange-correlation functional used was in PBE approximation,[78] and the norm-conserved pseudopotentials were based on Troullier-Martins parametrization.[79] All geometries were relaxed with forces smaller than 0.020 eV/Å. For reaction mechanism, we use nudged elastic band (NEB) method[80] as implemented in Atomic Simulation Environment (ASE) platform,[81] with 5 intermediate steps (images) and optimizations with forces smaller than 0.1 eV/Å.

**Acknowledgments**

The authors acknowledge the financial support from FAPESP (Grants Nos. 2012/50259-8; 2015/11779-4; 2018/25339-4), CNPq (Grant No. 408525/2018-5), CAPES (Grant No. 88887.310281/2018-00), and the Brazilian Nanocarbon Institute of Science and Technology (INCT/Nanocarbono). V.S.M. acknowledge FAPESP fellowship (Grant No. 2016/20799-1). L.S. also thanks the high-performance computer facilities provided by LoboC/NACAD/UFRJ. H.B.R. acknowledge FAPESP fellowships (Grants No. 2018/04926-9 and 2017/20100-0). A.R.C. acknowledges the FAPESP fellowship (Grant No. 2020/04374-6). We also acknowledge LNNano for the technical supporting during XPS analysis and AFM (Graphical abstract image).

# Supporting Information

**Table S1.** Peak position and FWHM of all Raman active BP modes for the spectra shown in Figure 1g.

| Mode | $A_g^1$ | | $B_{2g}$ | | $A_g^2$ | |
|---|---|---|---|---|---|---|
| | FWHM | Center | FWHM | Center | FWHM | Center |
| **Point 2** | 4.9 ± 0.1 | 363.5 ± 0.1 | 5.8 ± 0.7 | 440.6 ± 0.2 | 5.3 ± 0.2 | 468.1 ± 0.1 |
| **Point 1** | 5.0 ± 0.1 | 363.6 ± 0.1 | 4.9 ± 0.5 | 440.6 ± 0.4 | 5.2 ± 0.2 | 468.2 ± 0.1 |
| **No exposure** | 5.0 ± 0.1 | 363.3 ± 0.1 | 6.1 ± 1.5 | 441.0 ± 0.6 | 5.2 ± 0.1 | 468.5 ± 0.1 |

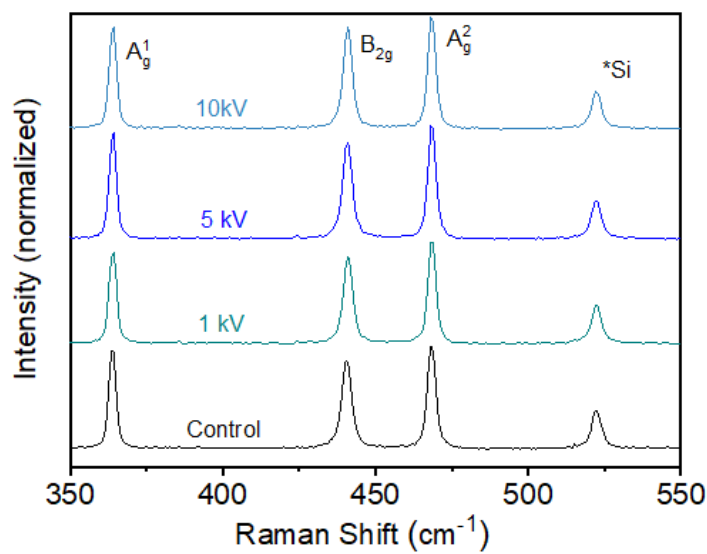

| Mode | $A_g^1$ | | $B_{2g}$ | | $A_g^2$ | |
|---|---|---|---|---|---|---|
| | FWHM | Center | FWHM | Center | FWHM | Center |
| **Control** | 2.3 ± 0.2 | 363.9 ± 0.1 | 3.4 ± 0.2 | 440.5 ± 0.1 | 2.8 ± 0.2 | 468.3 ± 0.1 |
| **1 kV** | 2.5 ± 0.2 | 363.9 ± 0.1 | 3.2 ± 0.3 | 441.0 ± 0.1 | 2.6 ± 0.2 | 468.5 ± 0.1 |
| **5 kV** | 2.4 ± 0.1 | 363.9 ± 0.1 | 3.3 ± 0.1 | 440.7 ± 0.1 | 2.7 ± 0.2 | 468.5 ± 0.1 |
| **10 kV** | 2.4 ± 0.1 | 364.1 ± 0.1 | 3.1 ± 0.1 | 440.8 ± 0.1 | 2.6 ± 0.1 | 468.5 ± 0.1 |

**Figure S1.** BP Raman characterization for different e-beam energy. Raman spectra of a BP flake exposed for 1 min at 7500x to different e-beam energy: 1 kV, 5 kV, and 10 kV. The table shows the peak position and FWHM of all Raman active BP modes for the spectra collected.

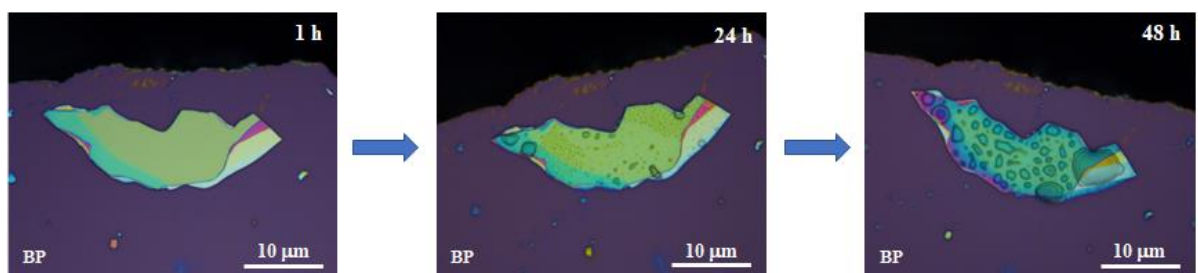

**Figure S2.** Monitoring of a BP flake non exposed to the e-beam. Optical microscope images for a BP-flake not imaged at the SEM after the exposure to ambient conditions (temperature: 19±1ºC and humidity: 60±5%) for different periods of time (hours).

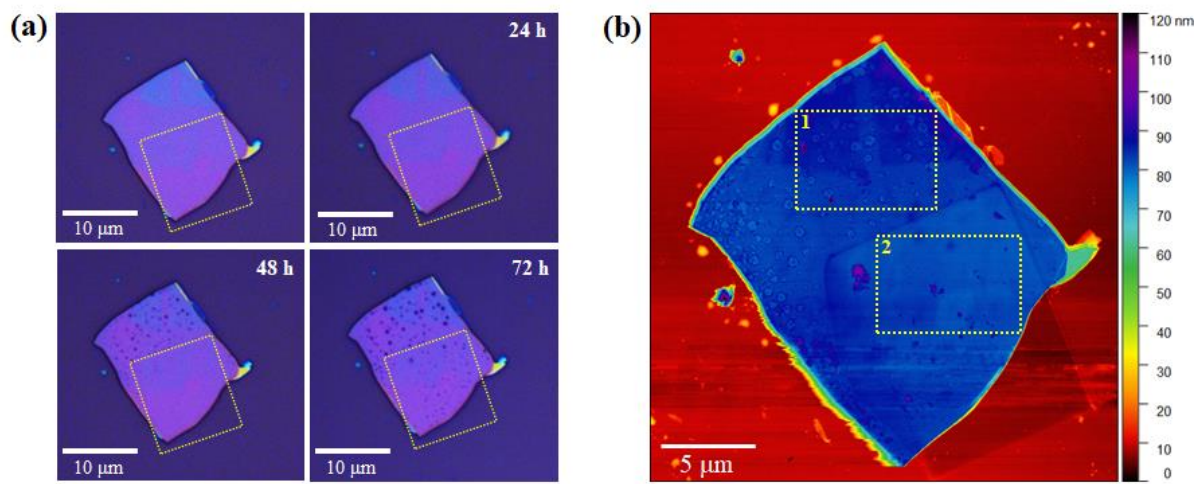

**Figure S3.** BP flake monitoring after e-beam exposure and N$_2$ healing. Optical microscope images for a BP-flake exposed to 5 kV after storage at ambient conditions (temperature: 19±1ºC and humidity: 60±5%) for different period of time. Highlighted region was exposed to the e-beam for longer period (5 kV, magnification of 10000x for 10 min). (b) AFM image for the same flake after 72 h showing RMS roughness of 2.26 nm and 1.18 nm for regions 1 and 2, respectively.

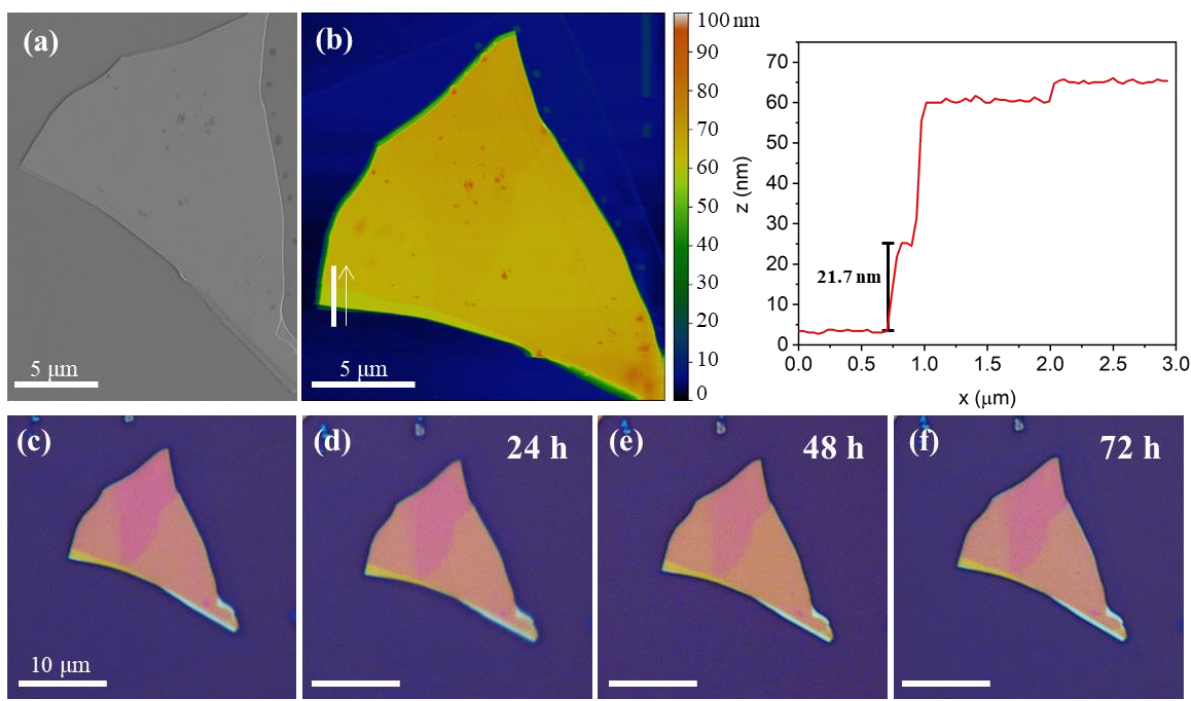

**Figure S4.** BP flake monitoring after e-beam exposure and N$_2$ healing. (a) SEM image of the flake exposed to 1 kV for 5 min (magnification 5000x). (b) AFM image for the same flake after 72 h storage at ambient conditions (temperature: 19±1ºC and humidity: 60±5%) and its respective

height profile. (c-f) Optical microscope images for the same flake immediatelly after the e-beam exposure and after 24, 48 and 78 hours at ambient conditions.

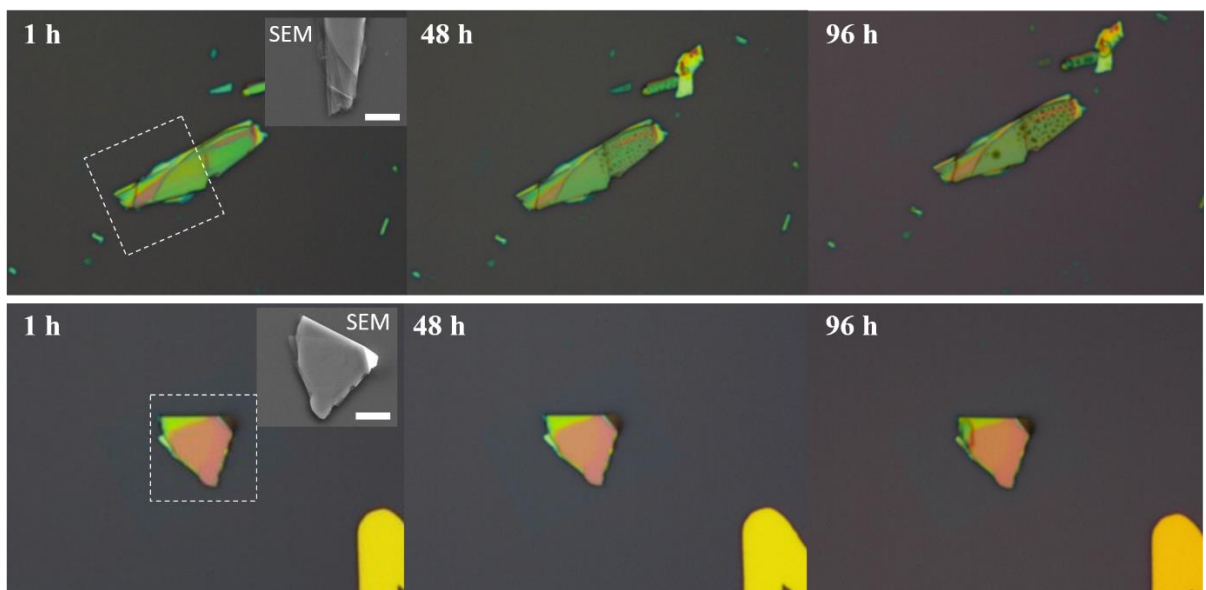

**Figure S5.** Temporal optical images of N-BP flakes exposed to 10 kV for 5 min (top panels) and 10 min (bottom panels). The insets bring the SEM image collected after the pre-selected time at 7500x. The scale bars are 5 µm. The samples were storaged at ambient conditions (temperature: 19±1ºC and humidity: 60±5%) for different period of time.

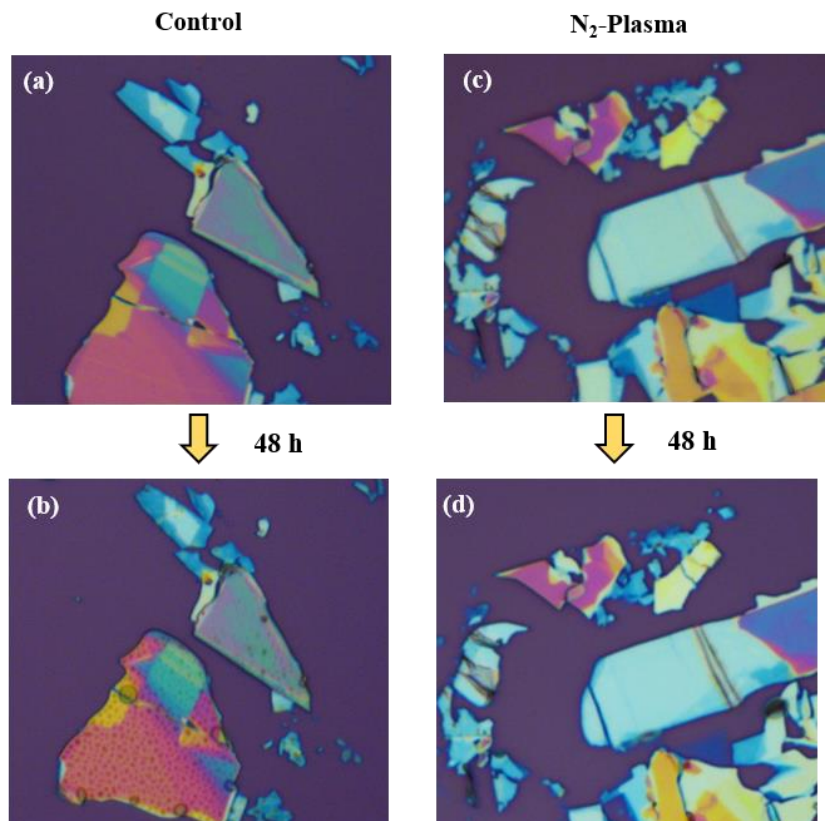

**Figure S6**. Optical microscopy images, taken after 1 and 48 hours in ambient conditions after sample preparation, for pristine BP (a-b), BP exposed to $N_2$ plasma for 5 min (c-d).

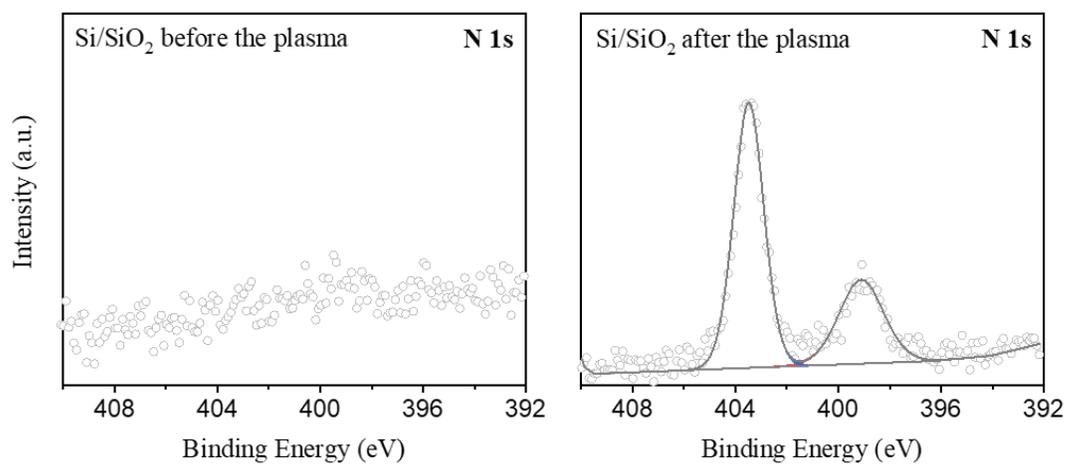

**Figure S7.** (a) High resolution N1s spectra of a Si/SiO₂ substrate before and after $N_2$-plasma (25 W, 5 min).

**Table S2.** Peak position and FWHM of the deconvoluted peaks for the High Resolution N1s and P2p XPS spectra shown in Figure 4a.

| Plasma $N_2$ | | Control | |
|---|---|---|---|
| *Position* | *FWHM* | *Position* | *FWHM* |
| 126.9 | 3.10 | 127.5 | 3.53 |
| 129.8 | 0.72 | 129.6 | 1.24 |
| 130.7 | 0.72 | 130.7 | 1.24 |
| 134.7 | 2.08 | 133.6 | 2.21 |
| 399.1 | 1.34 | 399.7 | 2.23 |
| 400.3 | 1.34 | - | - |
| 401.9 | 1.34 | - | - |
| 403.4 | 1.34 | - | - |

**Table S3.** Peak position and FWHM of all Raman active BP modes for the plasma treated samples (spectra shown in Figure 4b).

| Mode | $A_g^1$ | | $B_{2g}$ | | $A_g^2$ | |
|---|---|---|---|---|---|---|
| | FWHM | Center | FWHM | Center | FWHM | Center |
| **Control** | 4.7 ± 0.1 | 363.1 ± 0.1 | 5.5 ± 0.1 | 440.2 ± 0.1 | 5.1 ± 0.1 | 467.5 ± 0.1 |
| **Plasma Ar** | 4.7 ± 0.6 | 363.3 ± 0.2 | 5.8 ± 0.6 | 440.1 ± 0.2 | 5.1 ± 0.2 | 467.6 ± 0.1 |
| **Plasma $N_2$** | 4.8 ± 0.3 | 363.1 ± 0.1 | 5.7 ± 0.3 | 440.1 ± 0.1 | 5.3 ± 0.2 | 467.7 ± 0.5 |

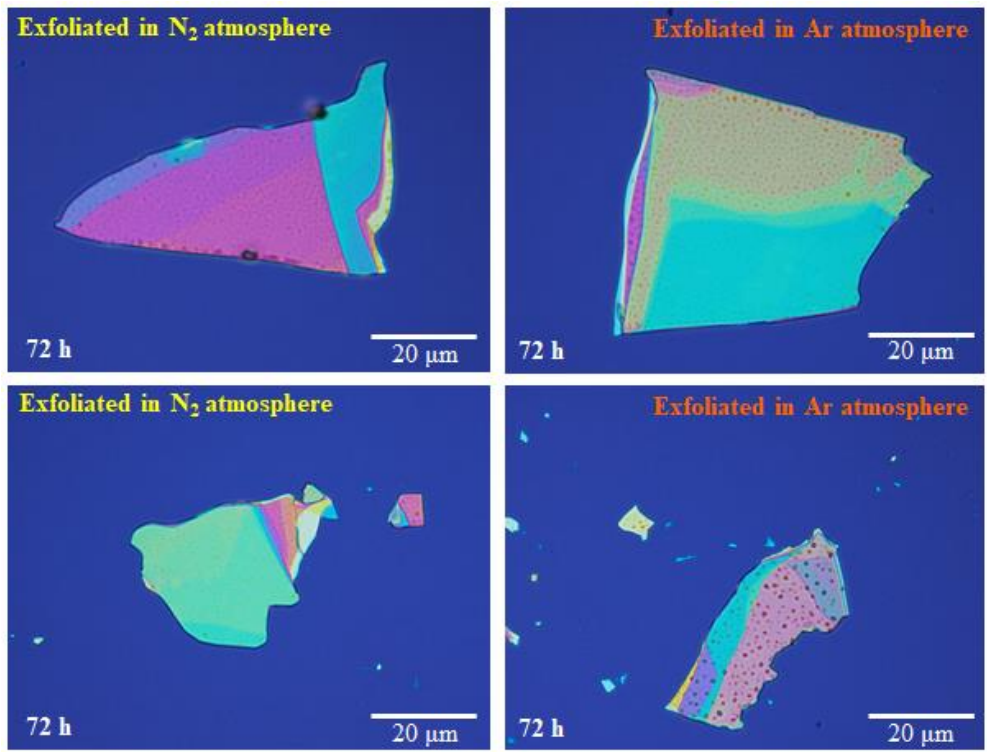

**Figure S8.** Monitoring of BP flakes exfoliated in different atmosphere. Optical images of BP flakes exfoliated in a $N_2$ atmosphere and Ar atmosphere after 72 h of exposition to the ambient conditions (temperature: 19±1°C and humidity: 60±5%).

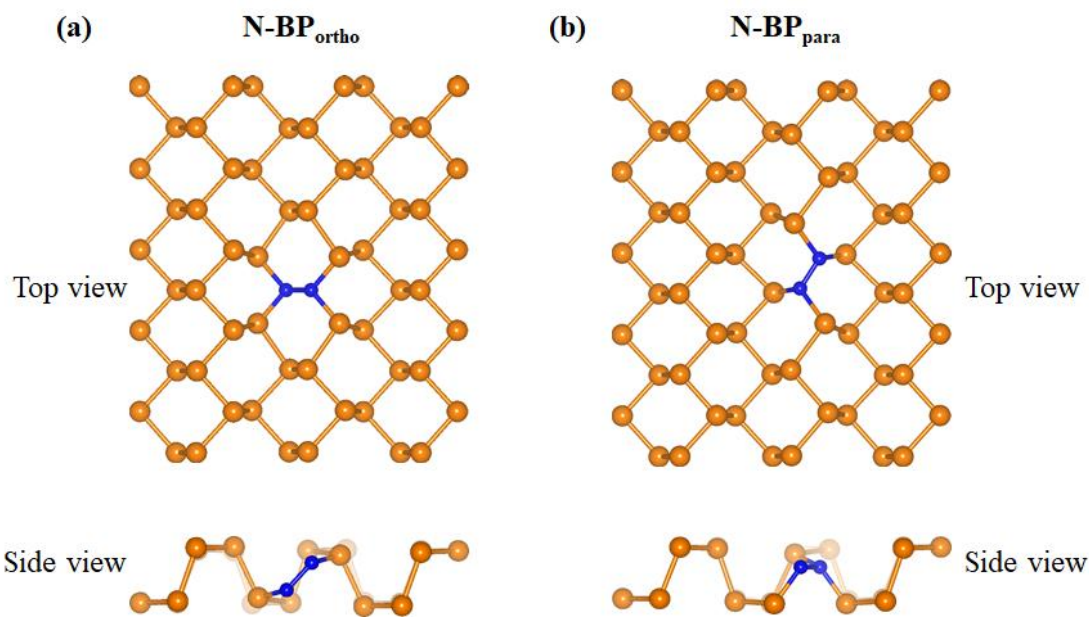

**Figure S9.** Side view and top view of (a) N-BP$_{ortho}$ and (b) N-BP$_{para}$.

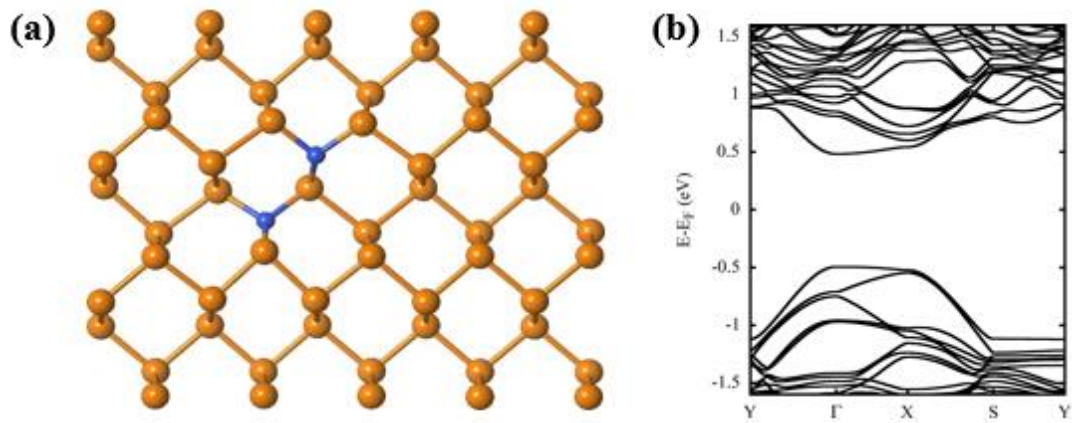

**Figure S10**. Top view of N-BP$_{NPN}$ after atomic restructuring and the (b) band structure for N-BP$_{NPN}$.

This is the authors´ version (post peer-review) of the manuscript:
V. Marangoni et. al. *Applied Surface Science*, 2021
(https://doi.org/10.1016/j.apsusc.2021.150450)**References**

[1] B. Li, C. Lai, G. Zeng, D. Huang, L. Qin, M. Zhang, M. Cheng, X. Liu, H. Yi, C. Zhou, F. Huang, S. Liu, Y. Fu, Black Phosphorus, a Rising Star 2D Nanomaterial in the Post-Graphene Era: Synthesis, Properties, Modifications, and Photocatalysis Applications, Small. 15 (2019) 1804565. https://doi.org/10.1002/smll.201804565.

[2] A. Castellanos-Gomez, Black Phosphorus: Narrow Gap, Wide Applications, J. Phys. Chem. Lett. 6 (2015) 4280–4291. https://doi.org/10.1021/acs.jpclett.5b01686.

[3] F. Xia, H. Wang, Y. Jia, Rediscovering black phosphorus as an anisotropic layered material for optoelectronics and electronics, Nat. Commun. 5 (2014) 4458. https://doi.org/10.1038/ncomms5458.

[4] H. Hedayat, A. Ceraso, G. Soavi, S. Akhavan, A. Cadore, C. Dallera, G. Cerullo, A.C. Ferrari, E. Carpene, Non-equilibrium band broadening, gap renormalization and band inversion in black phosphorus, 2D Mater. 8 (2021) 025020. https://doi.org/10.1088/2053-1583/abd89a.

[5] C. Chen, X. Lu, B. Deng, X. Chen, Q. Guo, C. Li, C. Ma, S. Yuan, E. Sung, K. Watanabe, T. Taniguchi, L. Yang, F. Xia, Widely tunable mid-infrared light emission in thin-film black phosphorus, Sci. Adv. 6 (2020) eaay6134. https://doi.org/10.1126/sciadv.aay6134.

[6] F. Xia, H. Wang, D. Xiao, M. Dubey, A. Ramasubramaniam, Two-dimensional material nanophotonics, Nat. Photonics. 8 (2014) 899–907. https://doi.org/10.1038/nphoton.2014.271.

[7] L. Viti, A. Politano, K. Zhang, M.S. Vitiello, Thermoelectric terahertz photodetectors based on selenium-doped black phosphorus flakes, Nanoscale. 11 (2019) 1995–2002. https://doi.org/10.1039/c8nr09060b.

[8] M.A. Huber, F. Mooshammer, M. Plankl, L. Viti, F. Sandner, L.Z. Kastner, T. Frank, J. Fabian, M.S. Vitiello, T.L. Cocker, R. Huber, Femtosecond photo-switching of interface polaritons in black phosphorus heterostructures, Nat. Nanotechnol. 12 (2016) 207–211. https://doi.org/10.1038/nnano.2016.261.

[9] X. Chen, J.S. Ponraj, D. Fan, H. Zhang, An overview of the optical properties and applications of black phosphorus, Nanoscale. 12 (2020) 3513–3534. https://doi.org/10.1039/C9NR09122J.

[10] L. Li, J. Kim, C. Jin, G.J. Ye, D.Y. Qiu, F.H. Da Jornada, Z. Shi, L. Chen, Z. Zhang, F. Yang, K. Watanabe, T. Taniguchi, W. Ren, S.G. Louie, X.H. Chen, Y. Zhang, F. Wang, Direct observation of the layer-dependent electronic structure in phosphorene, Nat. Nanotechnol. 12 (2017) 21–25. https://doi.org/10.1038/nnano.2016.171.

[11] H. Liu, A.T. Neal, Z. Zhu, Z. Luo, X. Xu, D. Tománek, P.D. Ye, Phosphorene: An unexplored 2D semiconductor with a high hole mobility, ACS Nano. 8 (2014) 4033–4041. https://doi.org/10.1021/nn501226z.

[12] S. Das, W. Zhang, M. Demarteau, A. Hoffmann, M. Dubey, A. Roelofs, Tunable transport gap in phosphorene, Nano Lett. 14 (2014) 5733–5739. https://doi.org/10.1021/nl5025535.

[13] J. Qiao, X. Kong, Z.-X. Hu, F. Yang, W. Ji, High-mobility transport anisotropy and linear dichroism in few-layer black phosphorus, Nat. Commun. 5 (2014) 4475. https://doi.org/10.1038/ncomms5475.

[14] D. He, Y. Wang, Y. Huang, Y. Shi, X. Wang, X. Duan, High-Performance Black Phosphorus Field-Effect Transistors with Long-Term Air Stability, Nano Lett. 19 (2019) 331–337. https://doi.org/10.1021/acs.nanolett.8b03940.

[15] Y. Liu, X. Duan, Y. Huang, X. Duan, Two-dimensional transistors beyond graphene and32

TMDCs, Chem. Soc. Rev. 47 (2018) 6388–6409. https://doi.org/10.1039/c8cs00318a.

[16] N.P. Rezende, A.R. Cadore, A.C. Gadelha, C.L. Pereira, V. Ornelas, K. Watanabe, T. Taniguchi, A.S. Ferlauto, A. Malachias, L.C. Campos, R.G. Lacerda, Probing the Electronic Properties of Monolayer MoS 2 via Interaction with Molecular Hydrogen, Adv. Electron. Mater. 5 (2019) 1800591. https://doi.org/10.1002/aelm.201800591.

[17] A. Castellanos-Gomez, L. Vicarelli, E. Prada, J.O. Island, K.L. Narasimha-Acharya, S.I. Blanter, D.J. Groenendijk, M. Buscema, G.A. Steele, J. V Alvarez, H.W. Zandbergen, J.J. Palacios, H.S.J. van der Zant, Isolation and characterization of few-layer black phosphorus, 2D Mater. 1 (2014) 025001. https://doi.org/10.1088/2053-1583/1/2/025001.

[18] J.O. Island, G.A. Steele, H.S.J. van der Zant, A. Castellanos-Gomez, Environmental instability of few-layer black phosphorus, 2D Mater. 2 (2015) 011002. https://doi.org/10.1088/2053-1583/2/1/011002.

[19] A. Ziletti, A. Carvalho, D.K. Campbell, D.F. Coker, A.H. Castro Neto, Oxygen Defects in Phosphorene, Phys. Rev. Lett. 114 (2015) 046801. https://doi.org/10.1103/PhysRevLett.114.046801.

[20] Y. Huang, J. Qiao, K. He, S. Bliznakov, E. Sutter, X. Chen, D. Luo, F. Meng, D. Su, J. Decker, W. Ji, R.S. Ruoff, P. Sutter, Interaction of Black Phosphorus with Oxygen and Water, Chem. Mater. 28 (2016) 8330–8339. https://doi.org/10.1021/acs.chemmater.6b03592.

[21] X. Liu, L. Xiao, J. Weng, Q. Xu, W. Li, C. Zhao, J. Xu, Y. Zhao, Regulating the reactivity of black phosphorus via protective chemistry, Sci. Adv. 6 (2020) eabb4359. https://doi.org/10.1126/sciadv.abb4359.

[22] J.D. Wood, S.A. Wells, D. Jariwala, K.-S. Chen, E. Cho, V.K. Sangwan, X. Liu, L.J. Lauhon, T.J. Marks, M.C. Hersam, Effective Passivation of Exfoliated Black Phosphorus Transistors against Ambient Degradation, Nano Lett. 14 (2014) 6964–6970. https://doi.org/10.1021/nl5032293.

[23] M. Caporali, M. Serrano-Ruiz, F. Telesio, S. Heun, A. Verdini, A. Cossaro, M. Dalmiglio, A. Goldoni, M. Peruzzini, Enhanced ambient stability of exfoliated black phosphorus by passivation with nickel nanoparticles, Nanotechnology. 31 (2020) 275708. https://doi.org/10.1088/1361-6528/ab851e.

[24] D. Grasseschi, D.A. Bahamon, F.C.B. Maia, A.H.C. Neto, R.O. Freitas, C.J.S. de Matos, Oxygen impact on the electronic and vibrational properties of black phosphorus probed by synchrotron infrared nanospectroscopy, 2D Mater. 4 (2017) 035028. https://doi.org/10.1088/2053-1583/aa8210.

[25] S. Kuriakose, T. Ahmed, S. Balendhran, V. Bansal, S. Sriram, M. Bhaskaran, S. Walia, Black phosphorus: Ambient degradation and strategies for protection, 2D Mater. 5 (2018) aab810. https://doi.org/10.1088/2053-1583/aab810.

[26] A. Favron, E. Gaufrès, F. Fossard, A.-L. Phaneuf-L'Heureux, N.Y.-W. Tang, P.L. Lévesque, A. Loiseau, R. Leonelli, S. Francoeur, R. Martel, Photooxidation and quantum confinement effects in exfoliated black phosphorus, Nat. Mater. 14 (2015) 826–832. https://doi.org/10.1038/nmat4299.

[27] G. Abellán, S. Wild, V. Lloret, N. Scheuschner, R. Gillen, U. Mundloch, J. Maultzsch, M. Varela, F. Hauke, A. Hirsch, Fundamental Insights into the Degradation and Stabilization of Thin Layer Black Phosphorus, J. Am. Chem. Soc. 139 (2017) 10432–10440. https://doi.org/10.1021/jacs.7b04971.

[28] Y. Abate, D. Akinwande, S. Gamage, H. Wang, M. Snure, N. Poudel, S.B. Cronin, Recent

Progress on Stability and Passivation of Black Phosphorus, Adv. Mater. 30 (2018) 1704749. https://doi.org/10.1002/adma.201704749.

[29] M. Druenen, Degradation of Black Phosphorus and Strategies to Enhance Its Ambient Lifetime, Adv. Mater. Interfaces. 7 (2020) 2001102. https://doi.org/10.1002/admi.202001102.

[30] J. Pei, X. Gai, J. Yang, X. Wang, Z. Yu, D.-Y. Choi, B. Luther-Davies, Y. Lu, Producing air-stable monolayers of phosphorene and their defect engineering, Nat. Commun. 7 (2016) 10450. https://doi.org/10.1038/ncomms10450.

[31] J.-S. Kim, Y. Liu, W. Zhu, S. Kim, D. Wu, L. Tao, A. Dodabalapur, K. Lai, D. Akinwande, Toward air-stable multilayer phosphorene thin-films and transistors, Sci. Rep. 5 (2015) 8989. https://doi.org/10.1038/srep08989.

[32] X. Chen, Y. Wu, Z. Wu, Y. Han, S. Xu, L. Wang, W. Ye, T. Han, Y. He, Y. Cai, N. Wang, High-quality sandwiched black phosphorus heterostructure and its quantum oscillations, Nat. Commun. 6 (2015) 7315. https://doi.org/10.1038/ncomms8315.

[33] R.A. Doganov, E.C.T. O'Farrell, S.P. Koenig, Y. Yeo, A. Ziletti, A. Carvalho, D.K. Campbell, D.F. Coker, K. Watanabe, T. Taniguchi, A.H.C. Neto, B. Özyilmaz, Transport properties of pristine few-layer black phosphorus by van der Waals passivation in an inert atmosphere, Nat. Commun. 6 (2015) 6647. https://doi.org/10.1038/ncomms7647.

[34] J. Kim, S.K. Baek, K.S. Kim, Y.J. Chang, E.J. Choi, Long-term stability study of graphene-passivated black phosphorus under air exposure, Curr. Appl. Phys. 16 (2016) 165–169. https://doi.org/10.1016/j.cap.2015.11.010.

[35] S. Gamage, A. Fali, N. Aghamiri, L. Yang, P.D. Ye, Y. Abate, Reliable passivation of black phosphorus by thin hybrid coating, Nanotechnology. 28 (2017) 265201. https://doi.org/10.1088/1361-6528/aa7532.

[36] Q. Li, Q. Zhou, X. Niu, Y. Zhao, Q. Chen, J. Wang, Covalent Functionalization of Black Phosphorus from First-Principles, J. Phys. Chem. Lett. 7 (2016) 4540–4546. https://doi.org/10.1021/acs.jpclett.6b02192.

[37] J.E.S. Fonsaca, S.H. Domingues, E.S. Orth, A.J.G. Zarbin, Air stable black phosphorous in polyaniline-based nanocomposite, Sci. Rep. 7 (2017) 1–9. https://doi.org/10.1038/s41598-017-10533-5.

[38] C.R. Ryder, J.D. Wood, S.A. Wells, Y. Yang, D. Jariwala, T.J. Marks, G.C. Schatz, M.C. Hersam, Covalent functionalization and passivation of exfoliated black phosphorus via aryl diazonium chemistry, Nat. Chem. 8 (2016) 597–602. https://doi.org/10.1038/nchem.2505.

[39] R. Gusmão, Z. Sofer, M. Pumera, Functional Protection of Exfoliated Black Phosphorus by Noncovalent Modification with Anthraquinone, ACS Nano. 12 (2018) 5666–5673. https://doi.org/10.1021/acsnano.8b01474.

[40] S. Walia, S. Balendhran, T. Ahmed, M. Singh, C. El-Badawi, M.D. Brennan, P. Weerathunge, M.N. Karim, F. Rahman, A. Rassell, J. Duckworth, R. Ramanathan, G.E. Collis, C.J. Lobo, M. Toth, J.C. Kotsakidis, B. Weber, M. Fuhrer, J.M. Dominguez-Vera, M.J.S. Spencer, I. Aharonovich, S. Sriram, M. Bhaskaran, V. Bansal, Ambient Protection of Few-Layer Black Phosphorus via Sequestration of Reactive Oxygen Species, Adv. Mater. 29 (2017) 1–8. https://doi.org/10.1002/adma.201700152.

[41] X. Tang, W. Liang, J. Zhao, Z. Li, M. Qiu, T. Fan, C.S. Luo, Y. Zhou, Y. Li, Z. Guo, D. Fan, H. Zhang, Fluorinated Phosphorene: Electrochemical Synthesis, Atomistic Fluorination, and Enhanced Stability, Small. 13 (2017) 1–10. https://doi.org/10.1002/smll.201702739.

This is the authors´ version (post peer-review) of the manuscript:
V. Marangoni et. al. *Applied Surface Science*, 2021
(https://doi.org/10.1016/j.apsusc.2021.150450)[42] C. Han, Z. Hu, L.C. Gomes, Y. Bao, A. Carvalho, S.J.R. Tan, B. Lei, D. Xiang, J. Wu, D. Qi, L. Wang, F. Huo, W. Huang, K.P. Loh, W. Chen, Surface Functionalization of Black Phosphorus via Potassium toward High-Performance Complementary Devices, Nano Lett. 17 (2017) 4122–4129. https://doi.org/10.1021/acs.nanolett.7b00903.

[43] J. Kim, S.S. Baik, S.H. Ryu, Y. Sohn, S. Park, B.G. Park, J. Denlinger, Y. Yi, H.J. Choi, K.S. Kim, Observation of tunable band gap and anisotropic Dirac semimetal state in black phosphorus, Science (80-. ). 349 (2015) 723–726. https://doi.org/10.1126/science.aaa6486.

[44] Y. Zhao, H. Wang, H. Huang, Q. Xiao, Y. Xu, Z. Guo, H. Xie, J. Shao, Z. Sun, W. Han, X.F. Yu, P. Li, P.K. Chu, Surface Coordination of Black Phosphorus for Robust Air and Water Stability, Angew. Chemie - Int. Ed. 55 (2016) 5003–5007. https://doi.org/10.1002/anie.201512038.

[45] Z. Yang, J. Hao, Recent Progress in Black-Phosphorus-Based Heterostructures for Device Applications, Small Methods. 2 (2018) 1–15. https://doi.org/10.1002/smtd.201700296.

[46] X. Zong, H. Hu, G. Ouyang, J. Wang, R. Shi, L. Zhang, Q. Zeng, C. Zhu, S. Chen, C. Cheng, B. Wang, H. Zhang, Z. Liu, W. Huang, T. Wang, L. Wang, X. Chen, Black phosphorus-based van der Waals heterostructures for mid-infrared light-emission applications, Light Sci. Appl. 9 (2020) 114. https://doi.org/10.1038/s41377-020-00356-x.

[47] X. Xuan, Z. Zhang, W. Guo, Doping-stabilized two-dimensional black phosphorus, Nanoscale. 10 (2018) 7898–7904. https://doi.org/10.1039/c8nr00445e.

[48] M. Ozhukil Valappil, M. Ahlawat, V.K. Pillai, S. Alwarappan, A single-step, electrochemical synthesis of nitrogen doped blue luminescent phosphorene quantum dots, Chem. Commun. 54 (2018) 11733–11736. https://doi.org/10.1039/c8cc07266c.

[49] C. Ji, C. Ji, A.A. Adeleke, L. Yang, B. Wan, H. Gou, Y. Yao, B. Li, Y. Meng, J.S. Smith, V.B. Prakapenka, W. Liu, G. Shen, W.L. Mao, H.K. Mao, Nitrogen in black phosphorus structure, Sci. Adv. 6 (2020) 1–8. https://doi.org/10.1126/sciadv.aba9206.

[50] N. Goyal, N. Kaushik, H. Jawa, S. Lodha, Enhanced stability and performance of few-layer black phosphorus transistors by electron beam irradiation, Nanoscale. 10 (2018) 11616–11623. https://doi.org/10.1039/C8NR01678J.

[51] H.B. Ribeiro, C.E.P. Villegas, D.A. Bahamon, D. Muraca, A.H. Castro Neto, E.A.T. de Souza, A.R. Rocha, M.A. Pimenta, C.J.S. de Matos, Edge phonons in black phosphorus, Nat. Commun. 7 (2016) 12191. https://doi.org/10.1038/ncomms12191.

[52] H.B. Ribeiro, M.A. Pimenta, C.J.S. de Matos, Raman spectroscopy in black phosphorus, J. Raman Spectrosc. 49 (2018) 76–90. https://doi.org/10.1002/jrs.5238.

[53] L.M. Malard, M.A. Pimenta, G. Dresselhaus, M.S. Dresselhaus, Raman spectroscopy in graphene, Phys. Rep. 473 (2009) 51–87. https://doi.org/10.1016/j.physrep.2009.02.003.

[54] F.M. Ardito, T.G. Mendes-De-Sá, A.R. Cadore, P.F. Gomes, D.L. Mafra, I.D. Barcelos, R.G. Lacerda, F. Iikawa, E. Granado, Damping of Landau levels in neutral graphene at low magnetic fields: A phonon Raman scattering study, Phys. Rev. B. 97 (2018) 1–6. https://doi.org/10.1103/PhysRevB.97.035419.

[55] H.B. Ribeiro, M.A. Pimenta, C.J.S. De Matos, R.L. Moreira, A.S. Rodin, J.D. Zapata, E.A.T. De Souza, A.H. Castro Neto, Unusual angular dependence of the Raman response in black phosphorus, ACS Nano. 9 (2015) 4270–4276. https://doi.org/10.1021/acsnano.5b00698.

[56] S. Gamage, Z. Li, V.S. Yakovlev, C. Lewis, H. Wang, S.B. Cronin, Y. Abate, Nanoscopy of Black Phosphorus Degradation, Adv. Mater. Interfaces. 3 (2016) 1–6. https://doi.org/10.1002/admi.201600121.
35

This is the authors´ version (post peer-review) of the manuscript:
V. Marangoni et. al. *Applied Surface Science*, 2021
(https://doi.org/10.1016/j.apsusc.2021.150450)[57]   S. Mignuzzi, A.J. Pollard, N. Bonini, B. Brennan, I.S. Gilmore, M.A. Pimenta, D. Richards, D. Roy, Effect of disorder on Raman scattering of single-layer MOS2, Phys. Rev. B. 91 (2015) 195411. https://doi.org/10.1103/PhysRevB.91.195411.

[58]   Z. Lin, B.R. Carvalho, E. Kahn, R. Lv, R. Rao, H. Terrones, M.A. Pimenta, M. Terrones, Defect engineering of two-dimensional transition metal dichalcogenides, 2D Mater. 3 (2016) 022002. https://doi.org/10.1088/2053-1583/3/2/022002.

[59]   J. Sonntag, J. Li, A. Plaud, A. Loiseau, J. Barjon, J.H. Edgar, C. Stampfer, Excellent electronic transport in heterostructures of graphene and monoisotopic boron-nitride grown at atmospheric pressure, 2D Mater. 7 (2020) 031009. https://doi.org/10.1088/2053-1583/ab89e5.

[60]   J. Gómez-Pérez, B. Barna, I.Y. Tóth, Z. Kónya, Á. Kukovecz, Quantitative Tracking of the Oxidation of Black Phosphorus in the Few-Layer Regime, ACS Omega. 3 (2018) 12482–12488. https://doi.org/10.1021/acsomega.8b01989.

[61]   F. Alsaffar, S. Alodan, A. Alrasheed, A. Alhussain, N. Alrubaiq, A. Abbas, M.R. Amer, Raman Sensitive Degradation and Etching Dynamics of Exfoliated Black Phosphorus, Sci. Rep. 7 (2017) 44540. https://doi.org/10.1038/srep44540.

[62]   W. Lu, H. Nan, J. Hong, Y. Chen, C. Zhu, Z. Liang, X. Ma, Z. Ni, C. Jin, Z. Zhang, Plasma-assisted fabrication of monolayer phosphorene and its Raman characterization, Nano Res. 7 (2014) 853–859. https://doi.org/10.1007/s12274-014-0446-7.

[63]   J. Jia, S.K. Jang, S. Lai, J. Xu, Y.J. Choi, J.-H. Park, S. Lee, Plasma-Treated Thickness-Controlled Two-Dimensional Black Phosphorus and Its Electronic Transport Properties, ACS Nano. 9 (2015) 8729–8736. https://doi.org/10.1021/acsnano.5b04265.

[64]   S. Kuriakose, T. Ahmed, S. Balendhran, G.E. Collis, V. Bansal, I. Aharonovich, S. Sriram, M. Bhaskaran, S. Walia, Effects of plasma-treatment on the electrical and optoelectronic properties of layered black phosphorus, Appl. Mater. Today. 12 (2018) 244–249. https://doi.org/10.1016/j.apmt.2018.06.001.

[65]   M. Van Druenen, F. Davitt, T. Collins, C. Glynn, C. O'Dwyer, J.D. Holmes, G. Collins, Evaluating the Surface Chemistry of Black Phosphorus during Ambient Degradation, Langmuir. 35 (2019) 2172–2178. https://doi.org/10.1021/acs.langmuir.8b04190.

[66]   X. Liu, Y. Bai, J. Xu, Q. Xu, L. Xiao, L. Sun, J. Weng, Y. Zhao, Robust Amphiphobic Few-Layer Black Phosphorus Nanosheet with Improved Stability, Adv. Sci. 6 (2019). https://doi.org/10.1002/advs.201901991.

[67]   N. Chauhan, V. Palaninathan, S. Raveendran, A.C. Poulose, Y. Nakajima, T. Hasumura, T. Uchida, T. Hanajiri, T. Maekawa, D.S. Kumar, N 2 -Plasma-Assisted One-Step Alignment and Patterning of Graphene Oxide on a SiO 2 /Si Substrate Via the Langmuir-Blodgett Technique, Adv. Mater. Interfaces. 2 (2015) 1400515. https://doi.org/10.1002/admi.201400515.

[68]   B. Wang, B.S. Kwak, B.C. Sales, J.B. Bates, Ionic conductivities and structure of lithium phosphorus oxynitride glasses, J. Non. Cryst. Solids. 183 (1995) 297–306. https://doi.org/10.1016/0022-3093(94)00665-2.

[69]   W. Hu, J. Yang, Defects in Phosphorene, J. Phys. Chem. C. 119 (2015) 20474–20480. https://doi.org/10.1021/acs.jpcc.5b06077.

[70]   J. V. Riffle, C. Flynn, B. St. Laurent, C.A. Ayotte, C.A. Caputo, S.M. Hollen, Impact of vacancies on electronic properties of black phosphorus probed by STM, J. Appl. Phys. 123 (2018) 044301. https://doi.org/10.1063/1.5016988.

[71]   A.A. Kistanov, Y. Cai, K. Zhou, S. V. Dmitriev, Y.-W. Zhang, The role of H 2 O and O 236

molecules and phosphorus vacancies in the structure instability of phosphorene, 2D Mater. 4 (2016) 015010. https://doi.org/10.1088/2053-1583/4/1/015010.

[72] V. Vierimaa, A. V Krasheninnikov, H.-P. Komsa, Phosphorene under electron beam: from monolayer to one-dimensional chains, Nanoscale. 8 (2016) 7949–7957. https://doi.org/10.1039/C6NR00179C.

[73] Y. Liu, F. Xu, Z. Zhang, E.S. Penev, B.I. Yakobson, Two-dimensional mono-elemental semiconductor with electronically inactive defects: The case of phosphorus, Nano Lett. 14 (2014) 6782–6786. https://doi.org/10.1021/nl5021393.

[74] R.G. Amorim, A. Fazzio, A. Antonelli, F.D. Novaes, A.J.R. Da Silva, Divacancies in graphene and carbon nanotubes, Nano Lett. 7 (2007) 2459–2462. https://doi.org/10.1021/nl071217v.

[75] L. Seixas, A. Carvalho, A.H. Castro Neto, Atomically thin dilute magnetism in Co-doped phosphorene, Phys. Rev. B. 91 (2015) 155138. https://doi.org/10.1103/PhysRevB.91.155138.

[76] W. Zheng, J. Lee, Z.W. Gao, Y. Li, S. Lin, S.P. Lau, L.Y.S. Lee, Laser-Assisted Ultrafast Exfoliation of Black Phosphorus in Liquid with Tunable Thickness for Li-Ion Batteries, Adv. Energy Mater. 10 (2020) 1–15. https://doi.org/10.1002/aenm.201903490.

[77] J.M. Soler, E. Artacho, J.D. Gale, A. García, J. Junquera, P. Ordejón, D. Sánchez-Portal, The SIESTA method for ab initio order- N materials simulation, J. Phys. Condens. Matter. 14 (2002) 2745–2779. https://doi.org/10.1088/0953-8984/14/11/302.

[78] J.P. Perdew, K. Burke, M. Ernzerhof, Generalized Gradient Approximation Made Simple, Phys. Rev. Lett. 77 (1996) 3865–3868.

[79] N. Troullier, J.L. Martins, Efficient pseudopotentials for plane-wave calculations, Phys. Rev. B. 43 (1991) 1993. https://doi.org/10.1103/PhysRevB.43.8861.

[80] G. Henkelman, H. Jónsson, Improved tangent estimate in the nudged elastic band method for finding minimum energy paths and saddle points, J. Chem. Phys. 113 (2000) 9978–9985. https://doi.org/10.1063/1.1323224.

[81] A. Hjorth Larsen, J. Jørgen Mortensen, J. Blomqvist, I.E. Castelli, R. Christensen, M. Dułak, J. Friis, M.N. Groves, B. Hammer, C. Hargus, E.D. Hermes, P.C. Jennings, P. Bjerre Jensen, J. Kermode, J.R. Kitchin, E. Leonhard Kolsbjerg, J. Kubal, K. Kaasbjerg, S. Lysgaard, J. Bergmann Maronsson, T. Maxson, T. Olsen, L. Pastewka, A. Peterson, C. Rostgaard, J. Schiøtz, O. Schütt, M. Strange, K.S. Thygesen, T. Vegge, L. Vilhelmsen, M. Walter, Z. Zeng, K.W. Jacobsen, The atomic simulation environment—a Python library for working with atoms, J. Phys. Condens. Matter. 29 (2017) 273002. https://doi.org/10.1088/1361-648X/aa680e.